\documentclass[prb,twocolumn,showpacs,amsmath,amssymb,superscriptaddress]{revtex4}

\usepackage{graphicx}
\usepackage{dcolumn}
\usepackage{bm}
\usepackage{color}

\begin{document}

\title{Evolution of the Fermi surface topology in doped 122 iron pnictides}

\author{Lihua Pan}
\affiliation{Texas Center for Superconductivity and Department of Physics, University of Houston, Houston, Texas 77204, USA}
\affiliation{School of Physics Science and Technology, Yangzhou University, Yangzhou 225002, China}

\author{Jian Li}
\affiliation{Texas Center for Superconductivity and Department of Physics, University of Houston, Houston, Texas 77204, USA}
\author{Yuan-Yen Tai}
\affiliation{Texas Center for Superconductivity and Department of Physics, University of Houston, Houston, Texas 77204, USA}
\author{Matthias J. Graf}
\affiliation{Theoretical Division, Los Alamos National Laboratory, Los Alamos, New Mexico 87545, USA}
\author{Jian-Xin Zhu}
\affiliation{Theoretical Division, Los Alamos National Laboratory, Los Alamos, New Mexico 87545, USA}
\affiliation{Center for Integrated Nanotechnologies, Los Alamos National Laboratory, Los Alamos, New Mexico 87545, USA}
\author{C. S. Ting}
\affiliation{Texas Center for Superconductivity and Department of Physics, University of Houston, Houston, Texas 77204, USA}

\begin{abstract}
Based on the minimum two-orbital model and the phase diagram recently proposed by Tai et al.\ (Europhys. Lett. \textbf{103}, 67001(2013)) for both electron- and hole-doped 122 iron-based superconducting compounds, we use the Bogoliubov-de Gennes equations to perform a comprehensive  investigation of  the evolution of the Fermi surface (FS) topology in the presence of the collinear spin-density-wave (SDW) order as the doping is changed. In the parent compound, the ground state is the SDW order, where the FS is not completely gapped, and two types of Dirac cones, one electron-doped and the other hole-doped emerge in the magnetic Brillouin zone. Our findings are qualitatively consistent with recent  angle-resolved photoemission spectroscopy and magneto-resistivity measurements.  We also examine the FS evolution of both electron- and hole-doped cases and compare them with measurements, as well as with those obtained by other model Hamiltonians.
\end{abstract}


\date{\today}


\maketitle

The discovery of the iron-based high-temperature superconductors has attracted intensive experimental and theoretical attention.  The parent compound (such as BaFe$_{2}$As$_{2}$ or 122 pnictide in short) is a bad metal with a collinear spin-density wave (SDW) order.  By doping either electrons or holes into the parent compound,   the SDW order becomes weakened and the superconductivity (SC) emerges.
Both phases appear to coexist and in some cases also compete with each other in certain doping regimes.~\cite{Pratt, Lester, Wang, Laplace, Avci, Rotter,HChen, Julien, Fernandes10}  As the compound is further doped, the SDW gets further suppressed.  Eventually, only the SC order prevails in the optimally and overdoped regimes. Many experiments~\cite{Christinason,CTChen} now suggest that the SC pairing symmetry in these compounds should be $s_{\pm}$-wave like~\cite{Mazin08} with the inter-band sign reversal of the pairing order, which can be simulated by a next-nearest-neighbor (NNN) pairing interaction.~\cite{GhamemiP}

Many microscopic Hamiltonians have been proposed to study the electronic band structure, the SDW and SC in the iron pnictides, ranging from tight-binding models that include all five Fe-3d orbitals~\cite{Five2} or even eight orbitals accounting for As-4p orbitals as well~\cite{Eight}, to three orbitals,~\cite{PALee08,Dagotto3} and down to the minimum models of two orbitals,~\cite{ZhangSC,Zhang, Hu,YYTai} or simply two bands,~\cite{Chubukovtwo,Chubukov2012,Chubukov09} as well as the low-energy effective model.~\cite{Effective} Each of these models has its own advantages and range of convenience for calculations.  For example, to study  certain specific properties of the 122 compounds, such as the Fermi surface (FS) evolution as a function of doping, quasiparticle excitations, superfluid density, and the local density  of states near an impurity or a magnetic vortex core,  the two-orbital models appear to have a clear numerical advantage, while retaining some of the orbital character of the low-energy bands.   Among the two-orbital ($d_{xz}$ and $d_{yz}$) models, the phenomenological approach of
~Zhang\cite{Zhang} takes into account the 2-Fe atoms per unit cell and  the asymmetry of the As atoms below and above of
the Fe plane, which breaks the C$_4$ symmetry. For this model the phase diagram of the electron-doped Ba(Fe$_{1-x}$Co$_{x}$As)$_{2}$ compounds has been  calculated,\cite{zhou1}  and the result is in qualitative agreement with experiments.~\cite{Pratt, Wang, Laplace, Julien, Christinason} The obtained FS evolution as a
function of electron doping~\cite{zhou1} and the FS at zero doping~\cite{zhou2} are consistent with the angle-resolved photoemission
spectroscopy (ARPES) experiments~\cite{KTerashima,YSekiba,Richard} and the electron- and hole-like Dirac cones, as observed indirectly  by
magneto-resitance measurements.~\cite{Huynh}  However,  this model failed to generate the experimental phase diagram for hole-doped
Ba$_{1-x}$K$_{x}$Fe$_{2}$As$_{2}$ compounds~\cite{Rotter,HChen} with the same set of parameters.  Thus the FS evolution of the hole-doped compounds has so far not been systematically studied in the literature. In a very recent work, Tai and co-workers~\cite{YYTai}  improved the original model in Ref.~\onlinecite{Zhang} to give a unified description of the entire phase diagram covering both the electron- and hole-doped regimes. To our knowledge, this is so far the only phenomenological 2-by-2-orbital model (2 Fe sites with 2 orbitals each), in which the resultant low-energy electronic dispersion agrees
qualitatively
well with density functional theory calculations of the electronic structure in the local density approximation (LDA) of the entire Brillouin zone (BZ) of the 122
compounds.~\cite{Five2,band1,band2,band3,band4} Notably, the obtained phase diagram
also agrees with the experimentally observed electron- and  hole-doped phase diagrams.~\cite{Pratt, Lester, Wang, Laplace, Avci, Rotter,HChen} In the present paper, we adopt this model to test further its validity by studying the FS topology of the hole-doped and electron-doped compounds.  At the same time we also compare the model results
with experiments and with previous theoretical studies.~\cite{zhou1,zhou2,Chubukov2012,YRan}	

We write the model Hamiltonian as\cite{YYTai}
\begin{equation}\label{hamiltonian}
H=H_{t}+H_{int}+H_{\Delta} ,
\end{equation}
where $H_{t}$ and $H_{int}$ are the single electron hopping and on-site electron-electron interaction terms, respectively. The kinetic energy term can be written as
$H_{t}=\sum_{\mathbf{i}\mu \mathbf{j}\nu\sigma}(t_{\mathbf{i}\mu \mathbf{j}\nu}c_{\mathbf{i}\mu\sigma}^\dag c_{\mathbf{j}\nu\sigma}+h.c.)-t_{0}\sum_{\mathbf{i}\mu\sigma}c_{\mathbf{i}\mu\sigma}^\dag c_{\mathbf{i}\mu\sigma}$,
and the electron-electron interaction term can be expressed in the mean-field approximation by
$H_{int}=U\sum_{\mathbf{i},\mu,\sigma\neq \bar{\sigma}}\langle n_{\mathbf{i}\mu\bar{\sigma}}\rangle n_{\mathbf{i}\mu\sigma}+
U^{'}\sum_{\mathbf{i},\mu\neq\nu,\sigma\neq \bar{\sigma}}\langle n_{\mathbf{i}\mu\bar{\sigma}}\rangle n_{\mathbf{i}\nu\sigma}
+(U^{'}-J_{H})\sum_{\mathbf{i},\mu\neq\nu,\sigma}\langle n_{\mathbf{i}\mu\sigma}\rangle n_{\mathbf{i}\nu\sigma}$,
where $\mathbf{i}$, $\mathbf{j}$ are lattice site indices; $\mu,\nu=1,2$ are the orbital indices for $d_{xz}$ and $d_{yz}$ orbitals;
$t_{0}$ is the chemical potential, which is determined by the electron filling per site $n$ with
$n_{\mathbf{i}\mu\sigma}=c_{\mathbf{i}\mu\sigma}^\dag c_{\mathbf{i}\nu\sigma}^{}$ and $U^{'}=U-2 J_{H}$.
At the mean field level, the pairing term is given by
$H_{\Delta}=\sum_{\mathbf{i}\mu \mathbf{j}\nu\sigma}(\Delta_{\mathbf{i}\mu \mathbf{j}\nu}c_{\mathbf{i}\mu\sigma}^\dag c_{\mathbf{j}\nu\bar{\sigma}}^\dag+h.c.)$.
Since this is a phenomenological Hamiltonian and dispersions are fit to low-energy LDA calculations or ARPES measurements, our classification in terms of d$_{xz}$ and d$_{yz}$ orbitals should not be taken literal, but rather as a convenient way to differentiate between the symmetries of the effective low-energy orbitals.
 A detailed symmetry analysis of our model has been presented in
Ref.~\onlinecite{YYTai}, in which the  C$_{4}$ symmetry breaking introduced. A more profound space group symmetry analysis of the iron pnictide superconductors can be found
in a recent paper.~\cite{Effective}

The Hamiltonian in Eq.~(\ref{hamiltonian}) is solved self-consistently through the multiorbital Bogoliubov-de Gennes equations in matrix notation,
\begin{equation}
\sum_{\mathbf{j}\nu}\left(
  \begin{array}{cc}
    H_{\mathbf{i}\mu \mathbf{j}\nu\uparrow} & \Delta_{\mathbf{i}\mu \mathbf{j}\nu}  \\
    \Delta_{\mathbf{i}\mu \mathbf{j}\nu}^{\ast}& -H_{\mathbf{i}\mu \mathbf{j}\nu\downarrow}^{\ast}  \\
  \end{array}
\right)
\left(
  \begin{array}{cc}
  \emph{u}_{\mathbf{j}\nu\uparrow}^{n}\\
  \emph{v}_{\mathbf{j}\nu\downarrow}^{n}\\
  \end{array}
  \right)
=E_{n}\left(
  \begin{array}{cc}
  \emph{u}_{\mathbf{i}\mu\uparrow}^{n}\\
  \emph{v}_{\mathbf{i}\mu\downarrow}^{n}\\
  \end{array}
  \right) ,
\end{equation}
in combination with the self-consistency equations for the electron density,
$n_{\mathbf{i}\mu}=\sum_{n}|\emph{u}_{\mathbf{i}\mu\uparrow}|^{2}f(E_{n})+\sum_{n}|\emph{v}_{\mathbf{i}\mu\downarrow}|^{2}[1-f(E_{n})]$,
and the SC order parameter,
$\Delta_{\mathbf{i}\mu \mathbf{j}\nu}=\frac{V_{\mathbf{i}\mu \mathbf{j}\nu}}{4}\sum_{n}
(\emph{u}_{\mathbf{i}\mu\uparrow}^{n}\emph{v}_{\mathbf{j}\nu\downarrow}^{n\ast}+
\emph{u}_{\mathbf{j}\nu\uparrow}^{n}\emph{v}_{\mathbf{i}\mu\uparrow}^{n\ast}){\rm tanh}(\frac{E_{n}}{2k_{B}T})$.
Here $f(E)$ is the Fermi-Dirac distribution function with Boltzmann constant $k_B$, $V_{\mathbf{i}\mu \mathbf{j}\nu}$ is the NNN pairing strength, with
$V_{\mathbf{i}\mu \mathbf{j}\mu}=V$, when $\mathbf{j}=\mathbf{i}\pm\hat{x'}\pm\hat{y'}$
and zero otherwise.

\begin{figure}
\includegraphics[width=0.45\textwidth,bb=15 20 234 150]{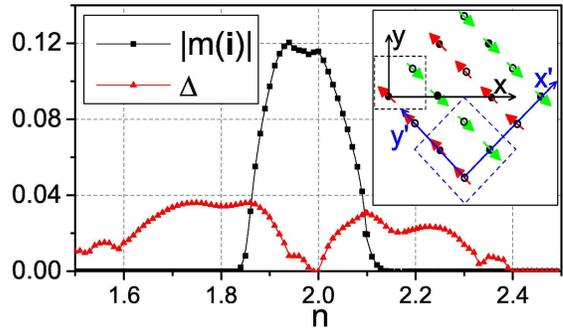}
\caption{\label{Fig. 1}(Color online) The phase diagram at zero temperature. The collinear SDW order parameter (black line with squares) and  the SC order parameter (red line with triangles) are shown. Inset: The schematic lattice structure of the Fe layer in the SDW state is plotted. Solid dots and circles denote nonequivalent Fe ions in different sublattices. Black and blue dashed squares denote the 2-Fe unit cell in the paramagnetic state and 4-Fe unit cell in the SDW state, respectively.}
\end{figure}

Throughout this work, we will use the six hopping parameters $t_{1-6}=(-1, 0.08, 1.35, -0.12, 0.09, 0.25 )$ and the three many-body interaction parameters $(U, J_H, V)=(3.2,0.6,1.05)$ from Ref.~\onlinecite{YYTai}. All energies are measured in units of $|t_1|$.
Before we give a detailed discussion about the effects of doping, we present the phase diagram of Tai's model in Fig.~\ref{Fig. 1}. In these calculations, the collinear SDW order parameter is defined as
$m(\mathbf{i})=(-1)^{\mathbf{i}_{x'}}\frac{1}{4}\sum_{\mu}(n_{\mathbf{i}\mu\uparrow}-n_{\mathbf{i}\mu\downarrow})$,
and the bulk SC order parameter is defined as the average over the lattice,
$\Delta=\frac{1}{8N}\sum_{\mathbf{i},\hat{\delta},\mu}\Delta_{\mathbf{i},\mathbf{i}+\hat{\delta},\mu}$.

\begin{figure}
\includegraphics[width=0.4\textwidth,bb=20 23 257 315]{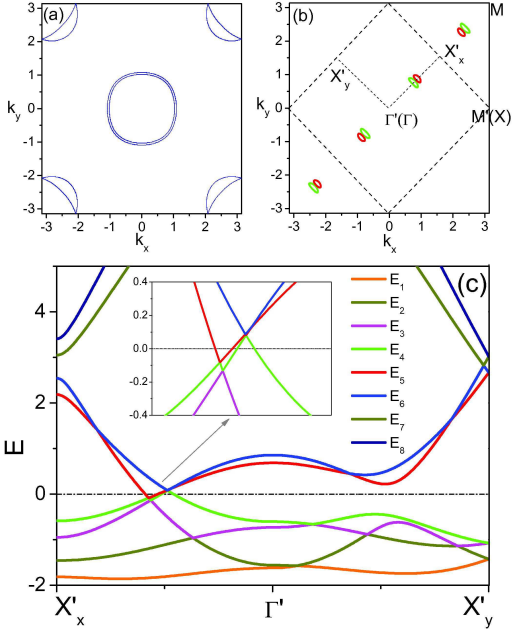}
\caption{\label{Fig. 2}(Color online)  (a) The normal state FS at half filling. (b) The ungapped magnetic FS at half-filling. The dash lines indicate the MBZ boundary in the SDW state. (c) The band structure along two directions $X_{\mbox{x}}^{'}-\Gamma^{'}$ and $\Gamma^{'}-X_{\mbox{y}}^{'}$. Inset: Enlarged view of the Dirac cones. }
\end{figure}

For the undoped case, the FS in the normal state contains two hole pockets around the $\Gamma=(0,0)$ point and two electron pockets around the $M=(\pi,\pi)$ point (Fig.~\ref{Fig. 2}(a)). Since the SDW order will enlarge the real-space unit cell, we choose the 4-Fe unit cell configuration as denoted by the blue dashed squares in the inset of Fig.~\ref{Fig. 1}, from which we can see that the antiferromagnetic order is along the $\mbox{x}'$ axis and the ferromagnetic order is along the $\mbox{y}'$ axis. We plot the zero-temperature magnetic FS of the undoped parent compound, obtained from our self-consistent calculation, in Fig.~\ref{Fig. 2}(b). Four small FS pockets appear in the magnetic Brillouin zone (MBZ) along the high-symmetry $\Gamma-M$ line, consistent with experiments.~\cite{Expsundoped1,Expsundoped2} We note that the pockets outside the MBZ are just replicas of those inside due to band folding in the SDW state. To reveal the nature of these FS pockets, we make a one-dimensional (1D) cut in Fig.~\ref{Fig. 2}(c) for the band structures along $X_{\mbox{x}}'-\Gamma'-X_{\mbox{y}}'$. We can see that two of the colored FS pockets (red) are electron-type and located around $(k_{x},k_{y})=\pm(0.287\pi,0.287\pi)$, while the other two (green) are hole-type and located around $(k_{x},k_{y})=\pm(0.244\pi,0.244\pi)$.
Along the line $X_{\mbox{x}}'-\Gamma'$, we predict the existence of two Dirac cones (one electron doped and the other hole doped). Indeed this is captured by recent ARPES
experiments~\cite{Richard} and is in good agreement with magneto-resistance measurements.~\cite{Huynh}
In retrospect, these calculations allow us to justify why  the present two-orbital model agrees so well with experiments.
This is because the density of states  due to the $d_{xy}$ orbital is practically zero at the Fermi energy in the presence of the SDW order~\cite{Shimojima} and thus may be neglected. Similar features for the parent compound have also been obtained by a different two-orbital
model.~\cite{zhou2}

The doping effect is expected to have an intimate impact on the FS with SDW order. In underdoped samples, where the SC and SDW orders coexist,  earlier work~\cite{Chubukov2012} investigated the effect of the SDW strength on the FS topology by keeping the chemical potential (or doping level) fixed. The FS with the SDW in the undoped case has also been shown to sensitively depend on the strength of  the onsite Coulomb interaction.~\cite{YRan} Therefore, it is necessary to choose a set of interaction parameters able to fit the phase diagram of the compounds for both electron- and hole-doped cases (see Fig.~1), and then to examine how the FS is changed  as the doping level varies. It appears that this issue has only been studied for the electron-doped case\cite{zhou1} based on the model of Ref.~\onlinecite{ZhangSC}. In the following, we first study the FS topology of the doped case by setting the SC order parameter $\Delta=0$. This is because the FS is determined by the SDW, not by SC. The effect of the SC is mainly to open a gap on parts of the FS pockets, where the SDW gap closes. Concretely,
in the optimally doped state, it can be predicted that $\Delta(\textbf{k})=2\Delta(cos(k_{x})+cos(k_{y}))$ in the BZ of the 2-Fe unit cell. In the coexistence region of SDW and SC, the reconstructed FSs are formed by mixing electron and hole bands but the SDW wave retains the gapped nature of the $s_{\pm}$-wave SC although the gap equation appears in a mathematically different way.~\cite{Chubukov2012,Chubukov09}
According to our numerical calculation,  the SDW gap is mainly affected by the doping  that destroying the nesting between the Fermi surfaces, not by the presence of the SC.  Further, the FS measured in the ARPES experiments for doped SC samples~\cite{KTerashima,YSekiba, ZXShen} usually did not exhibit the SC gaps.

\begin{figure}
\includegraphics[width=0.45\textwidth,bb=51 59 512 540]{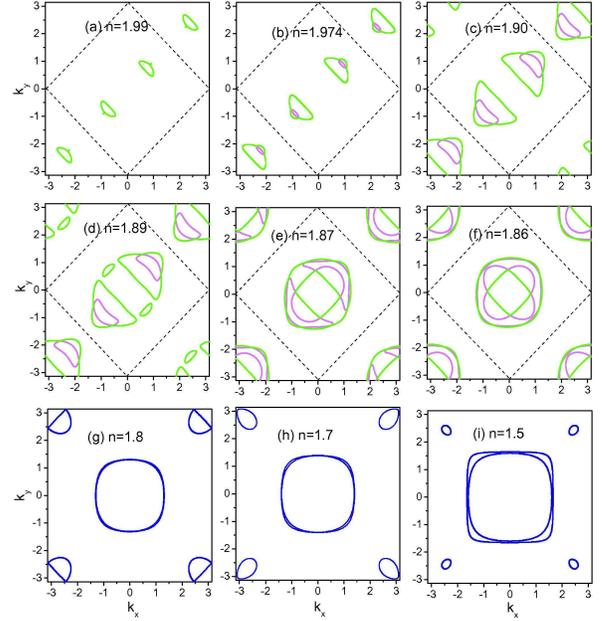}
\caption{\label{Fig. 3}(Color online) The doping evolution of the FS at the zero temperature on the hole-doped side of the phase diagram. The SC order is artificially set to zero in order to illustrate the effect of the SDW.  In panels (a)-(f), the green and magenta lines indicate the ungapped hole-type FS and the black dash lines indicate the MBZ boundary. The royal curves in (g)-(i) show the paramagnetic state FSs due to the totally depression of the SDW order.}
\end{figure}

When the iron pnictides are lightly hole-doped, i.e., away from half filling, the size of the hole pockets is enlarged, while that of the electron pockets is reduced. The electron pockets will then vanish completely at a small doping value as shown in Fig.~\ref{Fig. 3}(a). By further increasing doping, two new hole pockets appear in the same location where the electron pockets vanished (Fig.~\ref{Fig. 3}(b)-(c)). This can be easily seen from the inset of Fig.~\ref{Fig. 2}(c) by shifting the chemical potential downward. It is also worthwhile to point out that the band structure with the SDW depends strongly on the magnitude of the SDW order. When the doping level $\delta n\equiv $2-n$>0.1$, an additional pair of ungapped FS pockets appears in the diagonal $k_{x}=-k_{y}$ direction (Fig.~\ref{Fig. 3}(d)). The size of all these hole pockets is enlarged proportional to doping and then the FSs become closed around the $\Gamma$ point (two blue squares in Fig.~\ref{Fig. 3}(e)). Meanwhile, the magenta FS pockets are still located along the $k_{x}=k_{y}$ direction and stay gapped along the orthogonal direction, $k_{x}=-k_{y}$. When the system is even further doped, all the FSs become closed around the $\Gamma$ point and symmetric along $k_{x}=k_{y}$ and $k_{x}=-k_{y}$ directions as the SDW order becomes increasingly small (Fig.~\ref{Fig. 3}(f)).
So far we are not aware of any ARPES experiments in the hole-doped region with SDW.  Hence, our results will guide the search and interpretation of future experiments of the FS topology with SDW phase in the very underdoped regime.  In this regime, the SDW gap is large while the SC gap is comparatively small. If the ARPES experiment observes small gaps in certain k space-region while large gaps are detected in other part of the k space. Then the FS where the SC resides on could be easily determined. If one neglects the smaller SC gap, the FS with SDW as a function of doping should be experimentally obtained, and the results should be used to compare with our theoretical predictions.
 However, near the optimal doping, the SDW gap should be comparable to or much smaller than the SC gap. It would be hard for our theory to fit the experiments when the gaps of the SDW and SC orders have the same order of magnitude.

For samples with $n\leq1.86$, or $\delta n\geq0.14$, the SDW order disappears and thus no more band folding. We show the corresponding FSs in the 2-Fe BZ at $n=1.8$, $n=1.7$ and $n=1.5$ in Figs.~\ref{Fig. 3}(g)-(i), respectively. The four electronic pockets at the zone corner $M$ shrink, while the hole pockets at $\Gamma$ expand a little with increasing doping. It should be noticed that for extremely hole-doped samples, the small electron pockets near the $M$ points no longer touch the BZ boundary, which is consistent with available ARPES experiments,~\cite{DingH09,DingHJPCM} and so far has not yet been explained by other two-orbital models. In addition, Fig.~\ref{Fig. 3}(g) shows nearly degenerate hole-like FSs at the $\Gamma$ point, while experiments indicate two well separated hole like FSs.~\cite{DingH09,DingHJPCM,LWray2008,Dinget,PRichard,ZhouXJ} This is an intrinsic shortcoming of the two-orbital model, because
the outer hole pocket ($\beta$ FS sheet), which is missing in the present calculation, has a major $d_{xy}$ orbital component according to orbital-sensitive ARPES results.~\cite{DingH12} On the other side, the LDA calculations~\cite{ChiChengLee,VVildosola,CPLfirst} have shown that, although heavily hybridized, the main character of the bands that determine the FS are $d_{xz}$ and $d_{yz}$ orbitals, with small contributions of $d_{xy}$ at the hole pockets  and at most of the elongated portions of the electron pockets.
Even though the $d_{xy}$ orbital indeed affects the FS topology in the heavily hole-doped case, it has no qualitative impact on the collinear SDW and  SC orders, and other  thermodynamic  measurable quantities. In particular, we point out that although the two orbitals in our model are mainly of $d_{xz}$ and $d_{yz}$ character, some weight of $d_{xy}$-orbital character is also present.  This fact may qualitatively justify our minimum two-orbital model~\cite{YYTai}  as a phenomenological model for  122 pnictides.

\begin{figure}
\includegraphics[width=0.45\textwidth,bb=45 55 497 513]{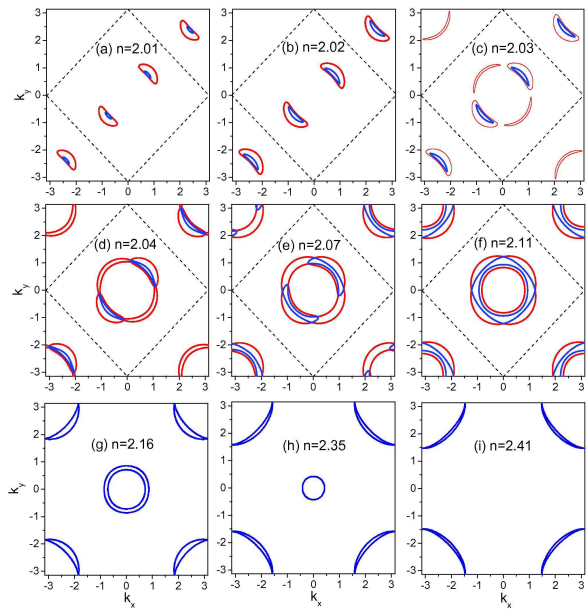}
\caption{\label{Fig. 4}(Color online) Similar to Fig. 3, but for the electron-doped side. The red and blue lines in panels (a)-(f) indicate the ungapped electronic FS.}
\end{figure}

In order to check the validity of the two-orbital model employed in the present work, we also examine the FS evolution with electron doping.
The results are shown in Fig.~\ref{Fig. 4}. In the lightly doped regime, the hole pockets are gradually getting smaller and finally disappear.
With further increased doping, two electron pockets appear (see Figs.~\ref{Fig. 4}(a) and (b)). This behavior is mirrored for hole doping and can be seen
from the inset of Fig.~\ref{Fig. 2}(c) by moving the chemical potential upward. However, in the intermediate doping regime around $n=2.03$, two new electron
FS arcs also appear along the diagonal $k_{x}=-k_{y}$ direction, see Fig.~\ref{Fig. 4}(c). This is quite different from the two-band result
in Fig.~2 of Ref.~\onlinecite{Chubukovtwo}, there the magnetic FS pockets are always along the $k_{x}=k_{y}$ direction. Recently, ARPES
experiments~\cite{ZXShen} were performed on strongly underdoped BaFe$_{2-x}$Co$_{x}$As$_{2}$ samples
in which the electronic structure of the detwinned crystal was observed. It can be seen that the anisotropic features in Fig.~\ref{Fig. 4}(c)
and ~\ref{Fig. 4}(d) are qualitatively comparable to the experimental observations as
shown in Fig.~1 of Ref.~\onlinecite{ZXShen}, if the crystal orientation across the twin boundaries is considered.
When the doping increases, the size of the electron FS pockets and arcs are enlarged,  and eventually
become large closed FSs around $\Gamma$, while the other pair of ungapped FS pockets remains along the $\Gamma$-$M$ line (Figs.~\ref{Fig. 4}(e)).
When the system is further doped, the inner blue electron pockets are also closed around the $\Gamma$ point (Fig.~\ref{Fig. 4}(f)).

For $n\geq 2.11$, the SDW order is totally suppressed, and the FS is then presented in the 2-Fe BZ, as shown in Fig.~\ref{Fig. 4}(g)-(i). It can be seen that in the heavily electron-doped sample, the hole-like FSs around the $\Gamma$ point become very small. First the inner one disappears, and finally both hole FSs disappear while the electron FS pockets at the zone corner become enlarged. A key finding of this work is that all these  model results are consistent with ARPES
experiments.~\cite{KTerashima,YSekiba}
In addition, the shape of the FS pockets around the $M$ point are round, while those obtained in previous work were more square.~\cite{zhou1}

\begin{figure}
\includegraphics[width=0.45\textwidth,bb=13 18 270 345]{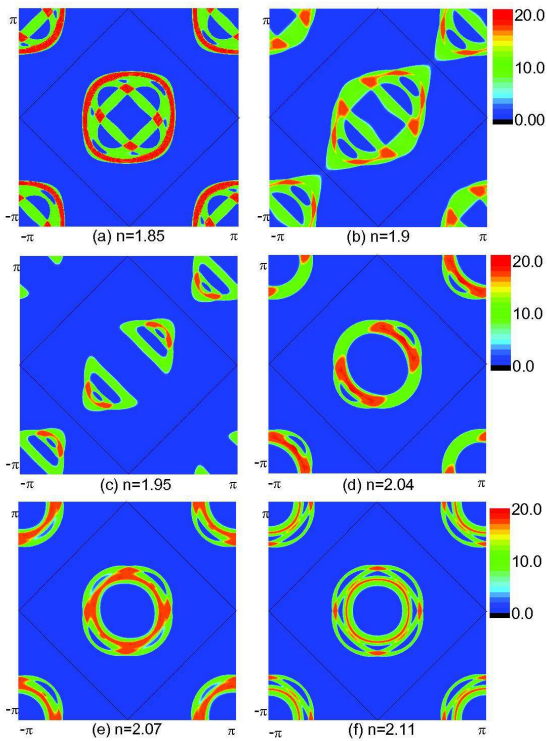}
\caption{\label{Fig. 5}(Color online) The spectra function $A(\textbf{k},\omega)$ integrated from $\omega=-0.1$ to $0.1$ at different doping levels: (a) $n=1.85$, (b) $n=1.9$, (c) $n=1.95$, (d) $n=2.04$, (e) $n=2.07$, and (f) $n=2.11$.}
\end{figure}

For a direct comparison to future experimental studies with varying doping levels, we calculate the spectral function $A(\textbf{k},\omega)=\sum_{\mathbf{i},\mu}A_{\mathbf{i},\mu}(\textbf{k},\omega)$, and integrate from $\omega=-0.1$ to $0.1$, which is proportional to the
photoemission intensity measured in ARPES experiments. The local and orbital-resolved spectral function is defined as
\begin{eqnarray}
\nonumber A_{\mathbf{i},\mu}(\textbf{k},\omega)=\sum_{n}[|\emph{u}_{\mathbf{i}\mu\uparrow}(\textbf{k})|^{2}\delta(E_{n}(\textbf{k})-\omega)+\\
|\emph{v}_{\mathbf{i}\mu\downarrow}(\textbf{k})|^{2}\delta(E_{n}(\textbf{k})+\omega)].
\end{eqnarray}
Our calculated spectral functions are shown in Fig.~\ref{Fig. 5}. It is known that when the sample is undoped or in the lightly hole- or electron-doped regimes, the spectral intensity is strong along the diagonal direction $k_{x}=k_{y}$ direction, but very weak along the other diagonal direction, $k_{x}=-k_{y}$.~\cite{Expsundoped1,Expsundoped2} When the electron or hole doping is increased, the intensity along $k_{x}=-k_{y}$  becomes enhanced (Figs.~\ref{Fig. 5}(b), ~\ref{Fig. 5}(d), and ~\ref{Fig. 5}(e)),
with obvious anisotropic characteristics induced by the existence of the SDW order. In the paramagnetic phase, as seen from Figs.~\ref{Fig. 5}(a) and ~\ref{Fig. 5}(f), the FSs become symmetric along the  $k_{x}=k_{y}$ and $k_{x}=-k_{y}$ directions.

The asymmetric characteristics of the magnetic FS along $k_x=k_y$ and $k_x=-k_y$ directions with the collinear SDW imply anisotropic transport and other properties in highly under-doped Fe-based superconductors.
Although the current ARPES experiments appear to be difficult to map out the evolution of the FS in the under-doped regime, the asymmetric effect of magnetic FS evolution should be easily reflected in other measurements, such as scanning tunneling microscopy (STM) experiments, which can measure the quasiparticle interference pattern and the local density of states  around a unitary impurity (Zn atom) or near a magnetic vortex.  All these signatures constitute subjects for future study.

In summary, we have studied for the first time systematically the FS evolution of the 122 parent compound as functions of  hole- and electron-doping.  At zero doping, there exist equal-sized electron-doped and hole-doped Dirac cones along the $\Gamma$-$M$ direction ($k_{x}=k_{y}$) in the BZ, i.e., the direction of the antiferromagnetic order. This is in good agreement with experiments.
When the 122 parent compound is lightly doped, the effect of doping is mainly to reduce the size of the Dirac cone-like pockets, while the SDW gap
closes up along the antiferromagnetic nesting direction. With further doping, additional parts of FSs  become ungapped along the orthogonal direction, $k_{x}=-k_{y}$. Then the SDW order is completely suppressed and the complete two-dimensional FSs appear in the heavily doped samples. We noticed that the FSs obtained for the heavily hole-doped regime seem  not to agree well with ARPES experiments, this is because the contribution from the $d_{xy}$ orbital is not adequately captured in the present study. However, our results with the SDW order can be used to guide future experiments on the evolution of the Fermi surface topology in very underdoped samples of Fe-pnictides, where the $d_{xy}$ orbital is greatly suppressed.   On the other hand,  we have also investigated the FS evolution as a function of electron doping.  All of our theoretical findings in this case are in qualitative agreement with experiments from under-doped to over-doped regime. This conclusion implies that the present model suits better for the electron-doped case than for the hole-doped case. We believe that the low energy physics of the 122 pnictides is originating mainly from the $d_{xz}$ and $d_{yz}$ orbitals, and further works are needed to support the validity of the present model.

\emph{Acknowledgments}. This work was supported in part by the Texas Center for Superconductivity at the University of Houston and by the Robert A. Welch Foundation under the Grant No. E-1146 (L.P., J.L., Y.-Y.T., \& C.S.T.). Work at Los Alamos was performed under the auspices of the U.S.\ DOE Contract No.~DE-AC52-06NA25396 through the LDRD program (Y.-Y.T.), the Office of Basic Energy Sciences (BES), Division of Materials Sciences \& Engineering (M.J.G.), and the  Center for Integrated Nanotechnologies, a BES user facility (J.-X.Z.). M.J.G. also thanks the Aspen Center for Physics for its hospitality, which is supported by the NSF under Grant No.~PHYS-1066293.


\end{document}